\documentclass{article}
\usepackage{spconf_one}
\usepackage{amsmath}
\usepackage{graphicx}

\usepackage{latexsym}
\usepackage{amsfonts}
\usepackage{graphicx} 
\usepackage{subfigure}

\usepackage{amssymb}
\usepackage{natbib}
\usepackage{enumerate}
\usepackage{amsmath}
\usepackage[usenames]{color}
\usepackage{longtable} 
\usepackage{tabularx}
\usepackage{chngpage}
\usepackage{lipsum}
\usepackage{booktabs}
\usepackage{verbatim}

\usepackage{bm}

\usepackage{subfigure}
\usepackage{tikz}
\usetikzlibrary{arrows}

\usepackage{slashbox}
\usepackage{color}
\usepackage{flushend}
\usepackage{multirow}
\usepackage{makecell}
\usepackage{color}
\usepackage{colortbl} 

\title{Parallel Metropolis chains with cooperative adaptation}
\name{L. Martino$^\Diamond$, V. Elvira$^{\dagger}$, D. Luengo$^{\ddagger}$, F. Louzada$^\Diamond$\thanks{This work has been partially supported by the Spanish government through the DISSECT (TEC2012-38058-C03-01) project, the BBVA Foundation through the MG-FIAR project, by the Grant 2014/23160-6 of S\~ao Paulo Research Foundation (FAPESP) and by the Grant 305361/2013-3 of National Council for Scientific and Technological Development (CNPq).}}

\address{$^\Diamond$ Institute of Mathematical Sciences and Computing, Universidade de S\~ao Paulo, S\~ao Carlos (Brazil). \\
$^{\dagger}$ Dep. of Signal Theory and Communic., Universidad Carlos III de Madrid,  Legan\'es (Spain).\\
$^{\ddagger}$ Dep. of Signal Theory and Communic., Universidad Polit\'ecnica de Madrid, Madrid (Spain).}
\begin{document}
%
\maketitle
\begin{abstract}
Monte Carlo methods, such as Markov chain Monte Carlo (MCMC) algorithms, have become very popular in signal processing over the last years. In this work, we introduce a novel MCMC scheme where parallel MCMC chains interact, adapting cooperatively the parameters of their proposal functions. Furthermore, the novel algorithm distributes the computational effort adaptively, rewarding the chains which are providing better performance and, possibly even stopping other ones. These extinct chains can be reactivated if the algorithm considers necessary. Numerical simulations shows the benefits of the novel scheme.  

\end{abstract}
\begin{keywords}
Interacting Parallel MCMC, Adaptive MCMC, cooperative adaptation.
\end{keywords}

\section{Introduction}
\label{sec:intro}

 Markov Chain Monte Carlo (MCMC) algorithms \citep{Robert04}  are widely employed in signal processing and communications  for Bayesian inference and optimization \citep{Doucet:MCSignalProcessing2005,Fitzgerald01,Jasra07b,Luengo13,MartinoIA2RMS15}. They draw random samples  from a complicated multidimensional target probability density function (pdf), $\pi({\bf x})$ with ${\bf x}\in\mathcal{D}\subseteq \mathbb{R}^n$, generating a Markov chain which converges to $\pi({\bf x})$. The performance, i.e., the speed of the convergence, depends strictly on the choice of a suitable proposal function $q({\bf x})$, and more specifically, on the discrepancy between $q({\bf x})$ and $\pi({\bf x})$.

 Speeding up the convergence has motivated an intense research activity. One active research line considers the design of an adaptive proposal density within MCMC techniques. Several schemes have been developed in order to tune online the parameters of the proposal density,  learning them from the previously generated samples
 \citep{Craiu09,Haario2001,Luengo13,MartinoIA2RMS15}.
On the other hand, the use of parallel chains instead of a single long chain has been studied for different reasons.  First of all, employing parallel chains allows the use of different proposal pdfs. Moreover, another important motivation is the interest in the implementation of MCMC techniques within a parallel architecture \citep{Calderhead14,Jacob11,OMCMC15}.
 Finally, a third reason is to speed up  the exploration of the state space \citep{Corander08,Jasra07b,O-MCMC,Smelly15,OMCMC15}. 
 Several works in the literature are focused on 
 producing an interaction among the different parallel chains \citep{Casarin13,Craiu09,O-MCMC,Smelly15,LAIS15}. The exchange of information among the chains can yield jointly two related benefits: produce a faster convergence of the chains to the target (e.g., reallocating ``lost'' chains  around a  mode of the target \citep{OMCMC15}), and help the cooperative exploration of the state space (for instance, generating a repulsion among the chains during a certain number of iterations \citep{Smelly15}). As a consequence,  the combined use of interacting parallel chains and adaptive proposal pdfs is considered of great interest in the literature \citep{Craiu09}.

In this work, we propose a novel scheme, called {\it parallel adaptive independent Metropolis} (PAIM), involving parallel MCMC chains which exchange information in order to adapt online their proposal densities. Namely, the interaction among the chains is carried out by a cooperative adaptation of the proposal functions. Each MCMC employs a proposal pdf, independent from the previous state of the chain. Each proposal is formed by a mixture of two densities, each one  determined by two parameters, a mean vector and covariance matrix. All the parameters are updated using empirical estimators (as in \citep{Haario2001,Luengo13}) applied to the  complete set or only to a subset of the previously generated states. The first component of each mixture aims to provide a global adaptation, the complete set of states is used. The second component of each mixture is adapted considering only  a subset of the previously generated states in order to learn local features of the target pdf. These subsets of states are built using a simple clustering-type strategy. This generates a cooperative adaptation, preventing that different proposals cover the same region, and allowing them to cover different modes of the target function, for instance. Furthermore, the novel algorithm is able to identify the proposal functions better located, and to allocate more computational effort into the corresponding chains. Indeed, PAIM also adapts the number of iterations of every chain, stopping those chains which are using a bad-located proposal. Moreover, PAIM is also able to turn back on certain chains when it is necessary.


%


%

%


\section{General setup}
\label{sec:problem}
In many different applications \citep{Corander08,Fitzgerald01,Jasra07b}, it is necessary to draw samples from a complicated $d$-dimensional target probability density function (pdf), $\bar{\pi}({\bf x})\propto \pi({\bf x})$, with ${\bf x} \in \mathcal{D} \subseteq \mathbb{R}^{d}$.
For this purpose, we consider the use of $N$ parallel Metropolis-Hastings chains \citep{Robert04}, each one employing a different independent proposal pdf $\psi_n({\bf x})$, defined as a mixture of two pdfs $q_{1,n}({\bf x})$ and $q_{2,n}({\bf x})$. More specifically, explicitly indicating also the parameters of the two components, we have 
$$
\psi_n({\bf x})=\frac{1}{2}q_{1,n}({\bf x}|{\bm \mu}_{1,n}, {\bf C}_{1,n})+\frac{1}{2}q_{2,n}({\bf x}|{\bm \mu}_{2,n}, {\bf C}_{2,n}),
$$
with ${\bf x}\in \mathcal{D}$, $n=1,\ldots,N$, where ${\bm \mu}_{i,n}$ represents a $d$-dimensional mean vector, and ${\bf C}_{i,n}$ is a $d\times d$ covariance matrix, for $i\in\{1,2\}$. 
In the novel method, the $N$ parallel chains interact in order to jointly all the parameters ${\bm \mu}_{i,n}$, ${\bf C}_{i,n}$, for $i=1,2$ and $\forall n$. The parameters  ${\bm \mu}_{1,n}$ and ${\bf C}_{1,n}$,  of the first component $q_{1,n}$ of every proposal $\psi_n$ are updated for providing a {\it global} adaptation, whereas the parameters ${\bm \mu}_{2,n}$ and ${\bf C}_{2,n}$ of $q_{2,n}$ are adapted in order to extract {\it local} features of the target pdf.  

 Furthermore, in the new scheme, each chain is performed for a different number of iterations, $K_n$, for $n=1,\ldots,N$. Indeed, the proposed algorithm is also able to determine online a subset of the $N$ chains which are obtaining the better performance. These chains are employed more often than others, namely increasing the total number of iterations $K_n$ performed by the corresponding chains. Let us denote the total number of desired samples as $L$, chosen in advance by the user. Then, we have $K_1+K_2\ldots +K_N=L$, i.e., PAIM is stopped when $L$ samples have been generated.\footnote{Note that we are including all the states in the ``burn-in'' periods of the $N$ chains. However, they can be discarded if desired.}
 

%

\section{Adaptive Parallel Independent Metropolis Algorithm}
\label{sec:algorithm}
\label{sec_Alg}

The {\it parallel adaptive independent Metropolis} (PAIM) algorithm works on two different time scales. First of all, the index $t$ denotes the current step of the algorithm, with 
\begin{equation}
t=0,\ldots, T_{tot},
\end{equation}
where the total number of steps $T_{tot}$ is not decided by the user, but automatically tuned by PAIM (we have always $T_{tot}\leq L$). Furthermore, we have have a different iteration index $k_n$ for each chain, such that 
\begin{equation}
k_n=0,\ldots,K_n, 
\end{equation}
for $n=1,\ldots,N$. The value of each $K_n$ is also decided by PAIM (recall that $\sum_{n=1}^N K_n=L$). At each step, the set  $\mathcal{A}_t$ contains the indices corresponding to the active chains. At each $t$-th step, every active chain performs one iteration (i.e., if the $j$-th chain is active, then $k_j=k_j+1$), whereas the inactive chains remain frozen.  For every $t\leq T_{train}$, where $T_{train}$ is chosen by the user, all the chains are active, i.e, 
$$
\mathcal{A}_t=\{1,2,\ldots,N\}, \qquad t\leq T_{train}.
$$
 Whereas, for $t>T_{train}$, we have $\mathcal{A}_t\subseteq\{1,2,\ldots,N\}$. The interacting adaptation is performed at any $t$ such that $ T_{train}<t<T_{stop}$. We consider the possibility of stopping the adaptation after $T_{stop}$ steps, since the adaptation could jeopardize the ergodicity of the chains. However, the numerical results described in Section \ref{SIMUSect} show that the algorithm seems to maintain the correct ergodicity properties.

The adaptation is performed as follows. During the first $T_{train}$ time steps the algorithm simply assigns the new current states generated at the $t$-th step,  to one chain among the $N$ possible, according to the minimum Euclidean distance between them and the means ${\bm \mu}_{2,n}^{(t)}$, for $n=1,...,N$. Thus, we allow the method to use a few iterations ($t=1,\ldots,T_{train}$) to collect information about the target, as in \citep{Haario2001}.
Afterwards, the algorithm adapts all the parameters ${\bm \mu}_{i,n}^{(t)}$ and ${\bf C}_{i,n}^{(t)}$, for $i\inÊ\{1,2\}$, and the set of active chains $\mathcal{A}_t$, until $t=T_{stop}$. On the one hand, the parameters ${\bm \mu}_{2,n}^{(t)}$ and ${\bf C}_{2,n}^{(t)}$ are updated using the empirical estimators for the means and covariances considering only the samples assigned to the $n$-th chain. On the other hand, the parameters ${\bm \mu}_{1,n}^{(t)}$ and ${\bf C}_{1,n}^{(t)}$ are adapted considered all the generated states so far. 
The chains are turned on or off taking into account the number of samples assigned to the $n$-th chain. 
PAIM is outlined in details below.
\newline
\newline
\newline
 {\bf1. Initialization:}   
 \begin{enumerate}
\item[1.1-] {\it Indices:} set $\ell=0$, $t=-1$,  $k_n=0$, for $n=1,\ldots,N$.
\item[1.2-] {\it Parameters:} choose the number of chains $N$ and the desired samples $L$. Select the positive values $\epsilon$, $T_{train}$, $T_{stop}$,\footnote{We recall that it is necessary to set $T_{train}<T_{stop}$.} the initial parameters ${\bm \mu}_{i,n}^{(0)}$, ${\bf C}_{i,n}^{(0)}$, for $i=1,2$, the initial states ${\bf x}_{n,0}$ and the counters $m_n=1$ for $n=1,\ldots,N$. Define the initial set of the active chains as $\mathcal{A}_0=\{1,2,\ldots,N\}$.

\end{enumerate}
 {\bf 2.  MH steps:} 
\begin{enumerate}
\item[2.1-] Set $t=t+1$, $\mathcal{Z}=\emptyset$ and $i=0$.
\item[2.2-] For all the active chains, i.e., for all the indices $j\in\mathcal{A}_t$: 
\begin{enumerate}
\item Sample ${\bf x}'$ from the $j$-th proposal pdf,	
 		$$
		 {\bf x}'\sim \psi_j^{(t)}({\bf x})=\frac{1}{2}\sum_{i=1}^2q_{i,j}^{(t)}({\bf x}|{\bm \mu}_{i,j}^{(t)}, {\bf C}_{i,j}^{(t)}).	
		$$
			\item Accept ${\bf x}_{j,k_j+1}={\bf x}'$ with probability
				\begin{equation}
					\small
						\alpha = \min\left[1, \frac{\pi({\bf x}') \psi_j^{(t)}({\bf x}_{j,k_j}) }{\pi({\bf x}_{j,k_j}) \psi_j^{(t)}({\bf x}')} \right].
				\label{AlphaEq}
				\end{equation}
			Otherwise, set ${\bf x}_{j,k_j+1}={\bf x}_{j,k_j}$. 
\item Set $\mathcal{Z}=\mathcal{Z} \cup \{ {\bf z}_{i+1}= {\bf x}_{j,k_j+1}\}$, and $i=i+1$.
\item Set ${\bm \theta}_{\ell+1}= {\bf x}_{j,k_j+1}$, $k_j=k_j+1$ and $\ell=\ell+1$.		
\item {\it Stop condition:} If $\ell>L$ then stop and return $\{{\bm \theta}_1,\ldots,{\bm \theta}_L\}$.
\end{enumerate}				
\end{enumerate}	
 {\bf3. Assignment} (if $t<T_{stop}$){\bf:} 
\begin{enumerate}
\item[3.1-] For all the vectors ${\bf z}_r\in \mathcal{Z}$, i.e., for $r=1,\ldots,|\mathcal{Z}|$ :
\begin{enumerate}
	\item Find the closest mean to ${\bf z}_i$ (w.r.t. the Euclidean distance), i.e., find the index
				\begin{equation}
        	n^*=\arg\min\limits_n || {\bm \mu}_{2,n}^{(t)}- {\bf z}_r ||.
        \end{equation}
        \item Set ${\bf s}_{n^*,m_{n^*}+1}={\bf z}_i$ and $m_{n^*}=m_{n^*}+1$. 
     \end{enumerate}
 \end{enumerate}
{\bf4. Adaptation} (if $T_{train}<t<T_{stop}$){\bf:}  
\begin{enumerate}
\item[4.1-] Set $\mathcal{A}_{t+1}=\emptyset$.
 \item[4.2-] For $n=1,\ldots,N$,  update the mean vectors,
  \begin{gather}
 \label{EqUp1}
 \begin{split}
           {\bm \mu}_{1,n}^{(t+1)}&=\widehat{{\bm \mu}}^{(t+1)}=\frac{1}{\ell}\sum_{i=1}^{\ell} {\bm \theta}_{i},  \\\quad 	{\bm \mu}_{2,n}^{(t+1)}&=\frac{1}{m_n}\sum_{i=1}^{m_n} {\bf s}_{n,i},
\end{split}
\end{gather}
update the covariance matrices,
\begin{gather}
 \label{EqUp2}
 \begin{split}
 {\bf C}_{1,n}^{(t+1)} &= \widehat{{\bf C}}^{(t+1)} = \frac{1}{\ell-1} \sum_{i=1}^{\ell} ({\bm \theta}_{i} -\widehat{{\bm \mu}})^2+\epsilon{\bf I}_d \\
 {\bf C}_{2,n}^{(t+1)}&=\frac{1}{m_n-1} \sum_{i=1}^{m_n} ({\bf s}_{n,i} -{\bm \mu}_{2,n}^{(t+1)})^2+\epsilon{\bf I}_d , 
\end{split}
\end{gather}
where ${\bf I}_d$ is $d\times d$ unit matrix. Moreover, if
         \begin{equation}
        \label{EqUp3}
	a_n = \left\lceil N\frac{m_n}{\sum_{j=1}^{N} m_j} \right\rceil >0,  
        \end{equation}          
then set $\mathcal{A}_{t+1}=\mathcal{A}_{t+1}\cup \{n\}$ (where $\lceil a \rceil$ denotes the smallest integer not smaller than $a$, with $a\in \mathbb{R}$). Otherwise, if $a_n=0$ the $n$-th chain is deactivated, i.e., not used at the next iteration. 
\end{enumerate}
{\bf 5. Repeat from step 2.} 
\section{Further considerations about PAIM}

Observe that the set $\mathcal{Z}$ contains all the new states generated in one specific step. This is refreshed at the beginning of every step $t$. Since the number of active chains is variable, the cardinality of $\mathcal{Z}$ is also changing with $t$. Moreover, we have denoted with ${\bf s}_{n,i}$ the $i$-th state assigned to the $n$-th chain. The counter $m_n$ indicates the number of states associated to the  $n$-th chain.
Note also that for all $t>T_{train}$ we have 
$$
{\bm \mu}_{1,n}^{(t)}=\widehat{{\bm \mu}}^{(t)}, \quad  {\bf C}_{1,n}^{(t)} = \widehat{{\bf C}}^{(t)}, \quad \mbox{ for } \ \ n=1,\ldots,N.
$$
Namely,  all the functions $q_{1,n}$ are updated using the empirical estimators of the mean and covariance of the target obtained from all the generated samples.  

It is important to remark that PAIM is able to distribute the computational efforts efficiently. Indeed, let us consider the initial use of a huge number $N$ of parallel chains, with proposal pdfs localized randomly over the state space. All the chains using a bad-localized proposal pdf would be quickly deactivated, i.e., only the chains with proposal pdfs located close to high probability regions would survive. However, the algorithm is also able to start up again certain chains if, after some steps, new states have been assigned to them. 
Finally note that, in the description of the algorithm, the parameters are updated using a block procedure, but efficient recursive update formulas can be employed (e.g., see \citep{Luengo13}), so that PAIM can be efficiently applied in high dimensional problems. 


\section{Numerical results}
\label{SIMUSect}
We consider a bi-dimensional ``banana-shaped'' target distribution \citep{Haario2001}, which is a benchmark function in the literature due to its nonlinear nature. Mathematically, it is  expressed  as 
$$
{\bar \pi}(x_1,x_2)\propto \exp\left(-\frac{1}{2\eta_1^2}\left(4-Bx_1-x_2^2\right)^2-\frac{x_1^2}{2\eta_2^2}-\frac{x_2^2}{2\eta_3^2}\right),
$$
where, we have set $B=10$, $\eta_1=4$, $\eta_2=5$, and $\eta_3=5$. The goal is to estimate the expected value $E[{\bf X}]$, where ${\bf X}=[X_1,X_2]\sim {\bar \pi}(x_1,x_2)$. We approximately compute the true value $E[{\bf X}]\approx[-0.4845,0]^{\top}$ using an exhaustive deterministic numerical method (with an extremely thin grid), in order to obtain the mean square error (MSE) of PAIM and the corresponding independent parallel MH chains with the same initial parameters but without adaptation. 

We consider $N\in\{5,10,50,100\}$ chains with Gaussian proposals $q_{i,n}({\bf x}|{\bm \mu}_{i,n}^{(0)}$, ${\bf C}_{i,n}^{(0)})$, $i=1,2$ and $n=1,\ldots,N$. The initial means as well as the initial states are chosen randomly at each run. More specifically, we have ${\bf x}_{n,0}\sim \mathcal{U}([-15,-15]\times[-15,15])$ and $ {\bm \mu}_{i,n}^{(0)} \sim \mathcal{U}([-15,-15]\times[-15,15])$. The initial covariance matrices are ${\bf C}_{i,n}^{(0)} = [\sigma^2 \quad 0; \ 0  \quad \sigma^2 ]^{\top}$ with $\sigma=10$. We consider $L=5000$ total number of samples. 
For PAIM, we test $T_{train}\in \{1,10,20\}$ and set  $\epsilon=0.4$, $T_{stop}=\infty$ (i.e., we never stop the adaptation).

\begin{figure*}[!tb]
\centering 
\centerline{
 \subfigure[]{\includegraphics[width=9cm]{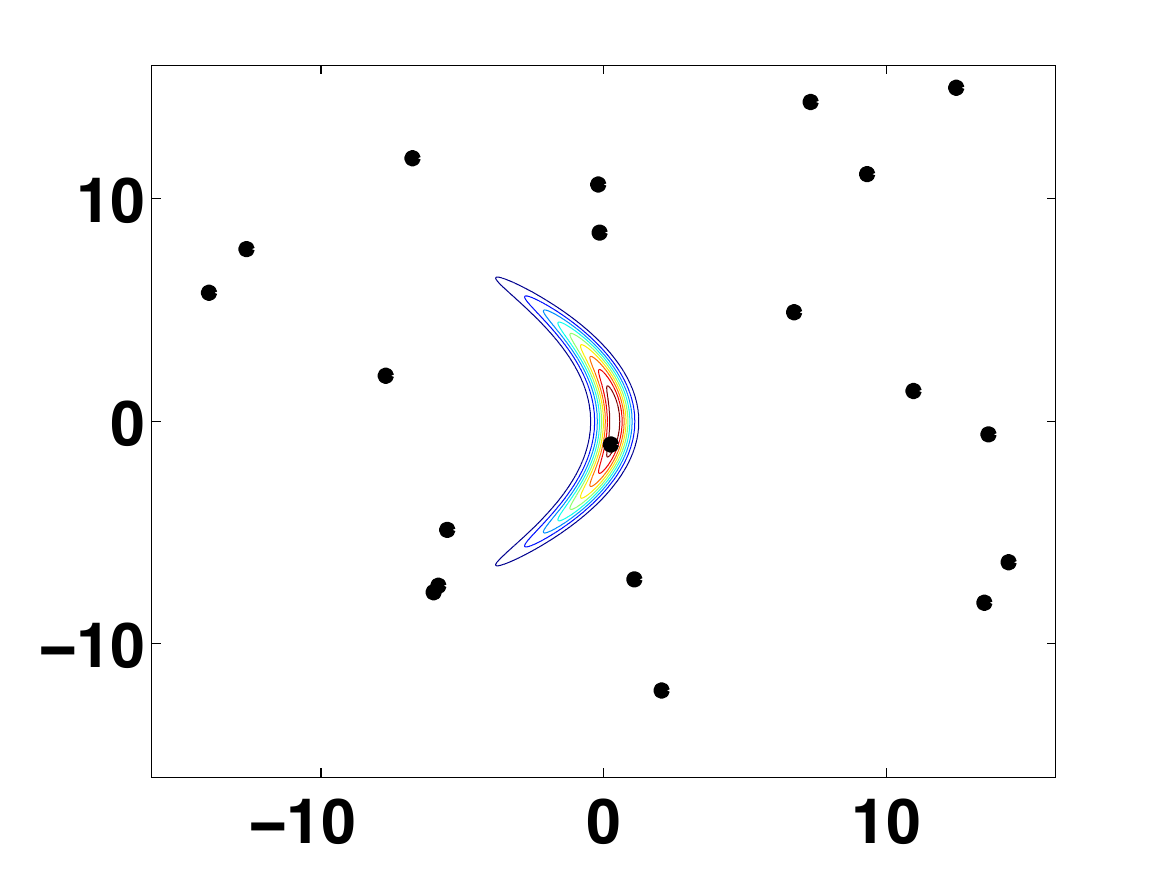}}
 \subfigure[]{ \includegraphics[width=9cm]{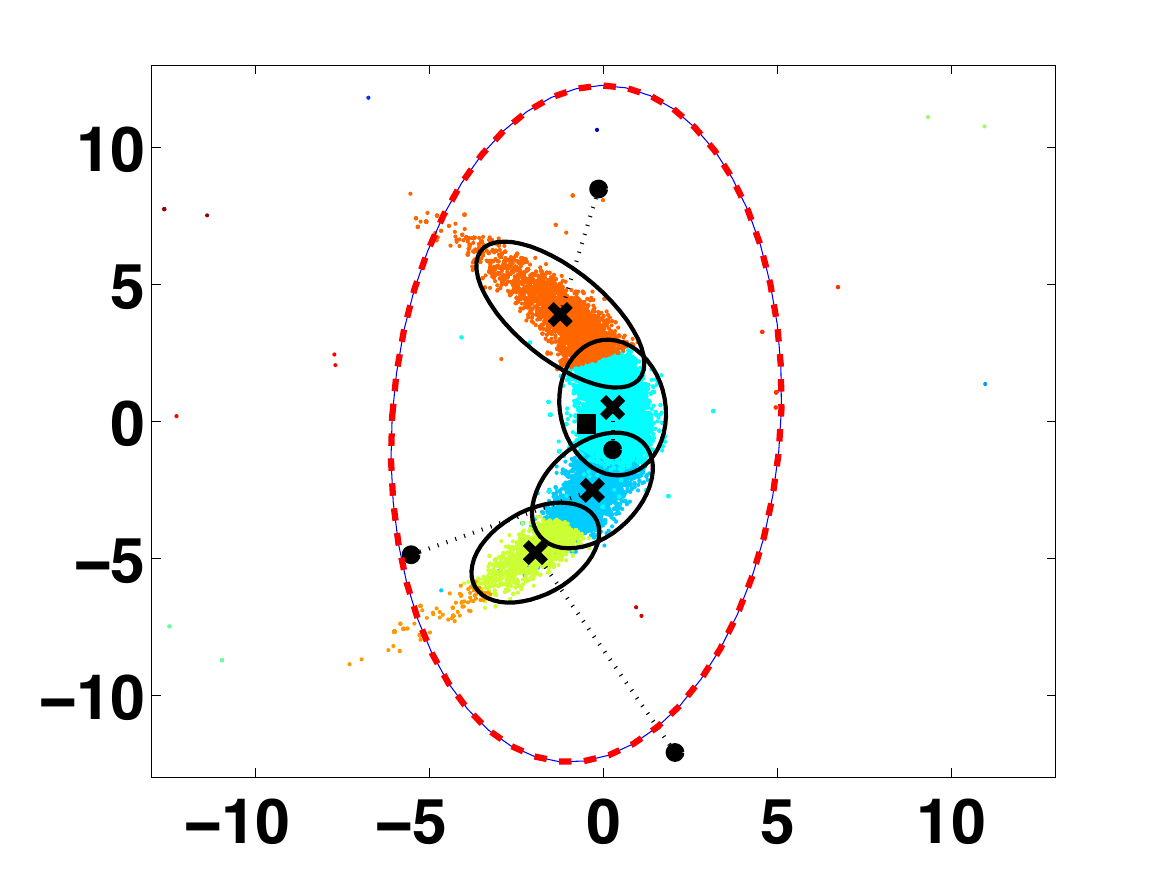}}
 }
  \caption{{\bf (a)} Contour plot of the banana-shaped target ${\bar \pi}$, and the initial means  ${\bm \mu}_{2,n}^{(0)}$ (circles) with $N=20$. {\bf (a)} We show the generated samples (dots), the initial (circles) and final means (x-marks) of the second component $q_{2,n}$ of the proposal of the final active chains. The final covariance ellipsoids are also depicted. The final mean of the first component of the proposals, ${\bm \mu}_{1,n}^{(T_{stop})}=\widehat{\bm \mu}^{(T_{stop})}$, is shown with a square jointly with the corresponding covariance ellipsoid with dashed line.   }
\label{fig1}
\end{figure*}

\begin{figure}[!tb]
\centering 
 \centerline{
\includegraphics[width=8cm]{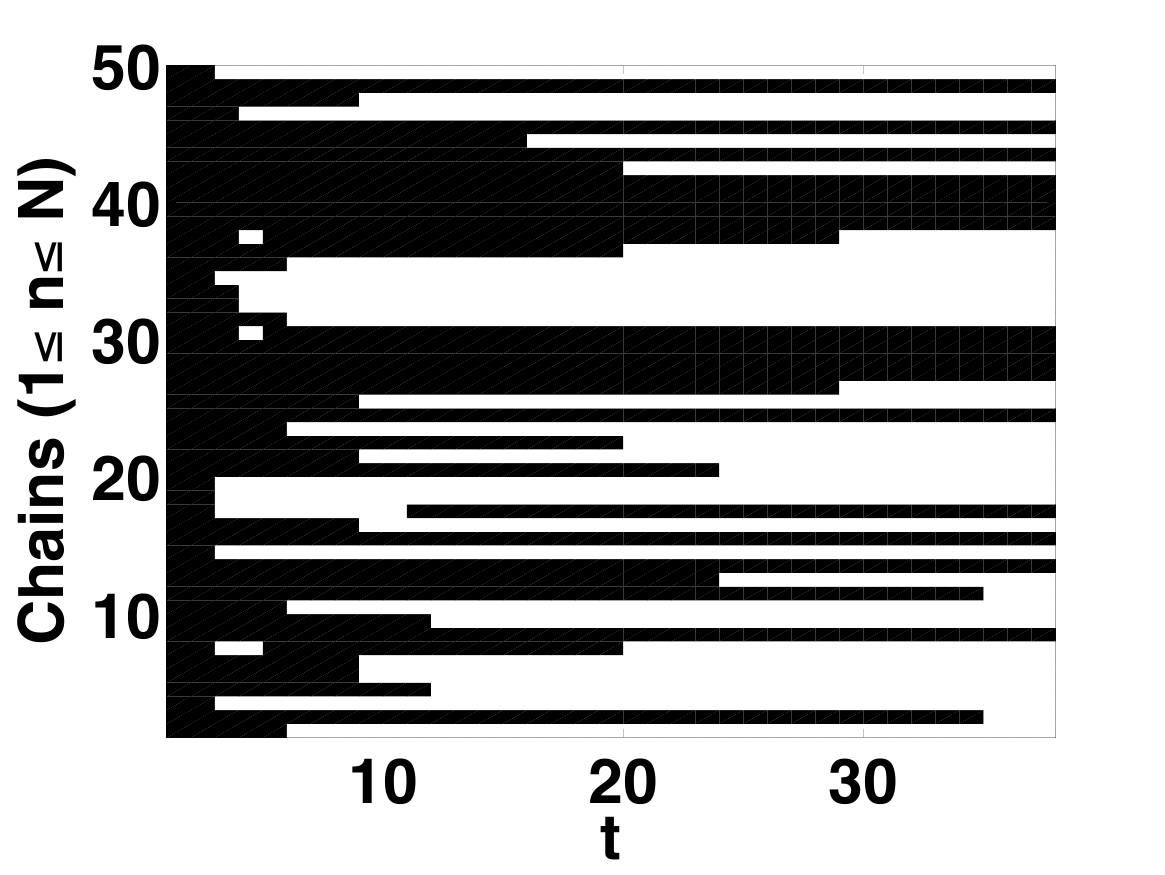}
 }
  \caption{ The indices of the active chains (denoted with black marks) at each step $t$ of PAIM  in one specific run, setting $N=50$, $T_{train}=2$, $L=1000$. In this run, the final number of active chains is $16$ and $T_{tot}=38$.  }
\label{fig2}
\end{figure}

The results are averaged $500$ over independent simulations, for each combination of parameters. PAIM always provides a smaller mean square error (MSE) in estimation of $E[{\bf X}]$ (averaging the two components) with respect to the corresponding independent parallel chains (IPCs).
 Table \ref{table1res} shows the percentage of reduction in  MSE obtained used PAIM with different $N$ and $T_{train}$. We can observe the advantage of the proposed adaptation procedure. Indeed, in this example, the minimum train period ($T_{train}=1$), provides the best results. 
\begin{table}[!htb]
\caption{Percentage of reduction in the MSE obtained using PAIM, with respect to IPCs. }
\label{table1res}
\footnotesize
\begin{center}
\begin{tabular}{|c|c|c|c|c|}
\hline
$T_{train}$ & $N=5$ & $N=10$ & $N=50$  & $N=100$    \\
\hline
\hline
1  & $51.34 \%$ & $58.05\%$ & $63.78\%$  &  $60.31\%$ \\
\hline
 10 & $46.95\%$ & $41.73\%$ & $44.23\%$ &  $35.65\%$  \\
 \hline
 20 & $29.81\%$ & $35.51\%$ & $33.29\%$ &  $22.33\%$  \\
\hline
\end{tabular}
\end{center}
\vspace{-0.5cm}
\end{table}%
Figure \ref{fig1}(a) shows the initial means of the second components of the proposals in PAIM, i.e., ${\bm \mu}_{2,n}^{(0)}$, and the contour plot of the target ${\bar{\pi}}$. Figure \ref{fig1}(b) depict the initial (circles) and final (x-marks) configurations of the means of the second components of the final active proposals, i.e., ${\bm \mu}_{2,n}^{(0)}$ and ${\bm \mu}_{2,n}^{(T_{stop})}$, obtained in one specific run (setting  $N=20$, $L=5 \ 10^4$). The final mean ${\bm \mu}_{1,n}^{(T_{stop})}=\widehat{\bm \mu}^{(T_{stop})}$ of the  first component common to all the proposals is depicted with a square. The figures also show with solid line the covariance ellipsoids, corresponding to the $\approx 90\%$ of the probability mass, of the second components    of the proposals  of the final active chains. The covariance ellipsoid corresponding to the first common component of the proposal pdfs is displayed with a dashed line.
Figure \ref{fig2} shows  the indices  corresponding to the active chains as function of $t$, in one specific run ($N=50$, $L=10^3$).  


\vspace{-0.2cm}
\section{Conclusions}
In this work, we have introduced a new interacting parallel MCMC scheme, where a cooperative adaptation of the proposal densities is performed. Furthermore, the computational effort is efficiently distributed by the the novel method among the set of parallel chains, according to their performance.

\bibliography{bibliografia}

\end{document}